\documentclass[twocolumn]{aa}
\usepackage{wasysym,txfonts,graphicx}
\newcommand{\lya}{Ly$\alpha$}
\newcommand{\Lya}{Ly$\alpha$}
\newcommand{\bet}{$\beta$}

\begin{document}

\title{HST/ACS Lyman $\alpha$ imaging of the {\em nearby} starburst 
ESO\,338-IG04\thanks{Based upon observations with the NASA/ESA 
{\em Hubble Space Telescope} obtained at the Space Telescope Science
Institute, which is operated by the Association of Universities for
Research in Astronomy, Inc., under NASA contract NAS5-26555}}

\author{M. Hayes
 \inst{1}
\and G. \"Ostlin
 \inst{1}
\and J. M. Mas-Hesse 
 \inst{2}
\and D. Kunth
 \inst{3}
\and C. Leitherer 
 \inst{4}
\and A. Petrosian 
 \inst{5}
}

\offprints{M. Hayes, \email{matthew@astro.su.se}}

\institute{Stockholm Observatory, AlbaNova University Centre, SE-106 91 Stockholm, Sweden
  \and Centro de Astrobiolog\'{\i}a (CSIC--INTA),
       E28850 Torrejon de Ardoz,
       Madrid, Spain
  \and Institut d'Astrophysique, Paris (IAP), 98 bis Boulevard Arago,
       F-75014 Paris, France
  \and Space Telescope Science Institute, 3700 San Martin Drive, Baltimore,
       MD 21218
  \and Byurakan Astrophysical Observatory and Isaac Newton Institute of 
       Chile, Armenian Branch, Byurakan 378433, Armenia
 }

\date{Received 17/01/2005 / Accepted //2005}

\abstract{
ESO\,338-IG04 (Tololo 1924-416) is a well known, luminous 
($M_V = -19.3$) Blue Compact Galaxy
in the local universe. It has a complex 
morphology indicating a recent merger and/or close interaction and contains
a central young starburst with compact star clusters of ages $\lesssim 40$Myr.
The galaxy was imaged using the 
Advanced Camera for Surveys (ACS) onboard the Hubble Space Telescope
(HST) in the Lyman $\alpha$ line and continuum. Using the Starburst99
synthetic spectra and other imaging data from the Wide Field and Planetary
Camera 2 we developed a technique that allows us to make 
the first photometrically valid subtraction of continuum from the \lya\ line.
The method allows us to disentangle the degenerate effects of age
and reddening by careful sampling of the UV continuum slope and 
4000\AA\ discontinuity.
Our results are in qualitative agreement with the models of \lya\ 
escape being regulated by kinematical properties of the interstellar 
medium.  
The line-only image shows \lya\ in both emission and absorption. Most 
notably, \lya\ emission is seen from central bright young clusters 
and is in spatial agreement
with the features present in a longslit spectrum 
taken with the Space Telescope Imaging Spectrograph.
The \lya\ is also seen in diffuse emission regions 
surrounding the central starburst where photons escape after one or more 
resonant scatterings in H{\sc i}. Quantitative photometry reveals a total flux
in the \lya\ line of
$f_{\rm Ly\alpha,TOT} = 194 \times 10^{-14}$erg~s$^{-1}$~cm$^{-2}$.
The \lya\ flux in a 10 $\times$ 20\arcsec\ elliptical 
aperture centred on the brightest central star cluster measures 
 $f_{\rm Ly\alpha,IUE} = 134\times 10^{-14}$ erg~s$^{-1}$~cm$^{-2}$ 
with an equivalent width of 22.6\AA. This is in close agreement with 
previous studies made using spectra from the IUE satellite to which our aperture
was created to match. Thus we demonstrate that we have software in place to create
line-only \lya\ maps of nearby galaxies. 
Analysis of 
parameter dependencies show our technique to be largely parameter 
independent, producing \lya\ maps indistinguishable from one another by eye 
and with \lya\ fluxes consistent with one another to better than 50\%. 
We see large
amounts of diffuse \lya\ emission that dominates the total \lya\ output 
which are interpreted as centrally produced \lya\ photons scattered by neutral 
hydrogen. By comparison of \lya\ fluxes with H$\alpha$ fluxes of a previous study, 
we estimate that each observed \lya\ photon has undergone 
$\gtrsim 2$ additional scatterings. 
We see that \lya\ line kinematics closely correlate with other kinematic 
tracers but, within this data, find no evidence for \lya\ emission 
or absorption from star clusters being a function of age.

 }

\titlerunning{\Lya\ imaging of ESO\,338-IG04}
\authorrunning{M. Hayes et al.}

\maketitle

 \section{Introduction}

The potential importance of the \lya\ emission line to probe galaxy formation
and evolution in the early universe has long since been recognised.
Partridge \& Peebles (1967) suggested the \Lya\ line as an important
spectral signature in young galaxies at high redshift since the expected
\Lya\ luminosity should amount to a few percent of the total galaxy
luminosity. If so, \Lya\ fluxes of order $10^{-15}$~erg~s$^{-1}$~cm$^{-2}$
should be detectable in galaxies forming stars with rates of
$10^2$~$M_{\odot}$~yr$^{-1}$ at redshifts ($z$) around 3.
Early observational attempts were largely disappointing and
demonstrated the scarcity of luminous \Lya\ emitters at high redshifts 
(Pritchet 1994). However, the situation has changed
dramatically in the recent years. Candidate
\lya \ emitters are now routinely detected at redshifts $z > 5$ using 
narrow-band imaging techniques, although the actual numbers of detected \Lya\ 
emitters at high redshift are smaller than was initially predicted 
(Frye, Broadhurst, \& Ben{\'{\i}}tez 2002; Fujita et al. 2003; 
Ouchi et al. 2003; Malhotra \& Rhoads 2002, Fynbo et al. 2001).

In principle, \lya \ can be used to quantitatively examine reionisation, 
star-formation rates, and correlation functions and grouping parameters in the 
distant universe. However, in order to proceed, it is essential for us to
understand the physics that regulates \lya \ escape. 
The situation is complicated by a number of factors; not least that \lya\ lies
in the far ultraviolet (UV) domain,
making it impossible to observe from 
the ground at redshifts less than $\sim 2$.  Furthermore, \lya \ is a 
resonant line and therefore is sensitive to
multiple scatterings and attenuation in
H{\sc i} clouds and galactic halos.
 This is evident from the \lya \ forest and damped \lya \
absorption seen in QSO absorption spectra. Multiple scatterings
can attenuate 
\lya \ via two mechanisms. Firstly, resonant scatterings redirect the \lya \
photons, thereby increasing the path length on which the \lya\ photons 
escape the neutral cloud with respect to the nearby continuum. 
This serves to increase the probability that the photons will be
destroyed by interactions with dust grains. Secondly, scattering of \lya\ 
photons by H{\sc i} can result not in one \lya\ photon but two photons of
lower energy. The probability of this two-photon emission is mildly 
temperature-sensitive but always $\sim 1/3$.

Early ultraviolet 
observations of a few nearby starbursts appeared to confirm the high-$z$
result: \Lya\ was surprisingly weak; weaker still in more dust-rich galaxies.
The reason was first thought to be multiple scatterings in H{\sc i} gas
increasing the probability of dust absorption (Neufeld 1991). However
subsequent Hubble Space Telescope (HST) observations showed that the problem
is much more complex. Kunth et al. (1998) discussed a set of eight
nearby star-forming galaxies observed
with the Goddard High Resolution Spectrograph (GHRS) on HST. Four
galaxies showed clear \Lya\ P-Cygni profiles indicating large-scale outflows
of the interstellar medium (ISM). In contrast, the four other
galaxies show damped \Lya\ absorption, regardless of
their dust content. Thuan \& Izotov (1997) studied the unique starburst
Tol1214-277 whose very strong and symmetric \Lya\ line shows no trace of
blueshifted absorption.
A subset of the analyzed starburst galaxies were observed with STIS 
(Mas-Hesse et al. 2003), in order to understand the kinematical structure 
of the outflowing gas. The main observational result of this work was the identification 
of rather large expanding shells of neutral gas, spanning up to several
 kiloparsecs and covering  a much larger area than the starburst itself. 

The complex nature of the \lya \ escape probability revealed by the GHRS 
and STIS spectroscopy (Kunth et al. 1998, Mas-Hesse et al. 2003) raised 
additional issues. In the case where
the UV source is shielded by a slab of static neutral hydrogen, the galaxy
appears as a damped \lya \ absorber. However, if the dust content is small
the \lya \ photons may, after multiple resonant scatterings, diffuse out 
over a larger area (although attenuated by the two-photon process).
This would create a bias in the spectroscopic studies
which usually target the regions of peak UV intensity - under this scenario
the places where we do {\em not} expect to see \lya \ in emission. Another
possibility is that the UV-continuum sources are partly shielded by a clumpy
medium, in which case we would see a mixture of absorption and emission. 
In cases where the ISM has a non-zero radial velocity with respect to the
UV continuum source, \lya \ will appear in emission with a characteristic
P-Cygni profile. Since radial velocity of the ISM is so important in 
determining \lya \ escape, morphology and kinematics of the galaxy play a 
vital role as certain kinematic configurations may enhance the escape 
probability. This in turn implies that the \lya \ escape probability can
vary greatly across the starburst region and, for example, during a merger
the escape probability may be enhanced. ESO\,338-IG04 is a perfect galaxy to 
study in this respect because it exhibits \lya\ that has escaped via all
of the mechanisms listed.

In the distant universe, \lya \ imaging and low resolution spectroscopic
techniques are now successfully used to find large numbers of
\lya \ emitting galaxies.
However, without a proper understanding of the \lya \ emission processes
this line cannot be used to estimate star formation rates and \lya\ becomes
dubious to use to study clustering if the biases are not properly known. 
If the star forming activity of a high redshift galaxy is connected with 
its environment, e.g. induced by merger or interaction, the \lya \ escape 
probability will not be independent of this parameter.

These considerations led us to start a pilot program to image {\em local} 
starburst galaxies in the \lya \ line using the solar blind channel (SBC)
of the Advanced Camera for Surveys (ACS) on HST; thereby allowing us to
study the \lya \ emission and absorption morphology in detail. 
A sample of six galaxies with a range of luminosities ($M_V = -15$ to $-21$) 
and metallicities ($0.04 Z_\odot {\rm ~to} \sim Z_\odot$) were selected, 
including previously known \lya \ emitters as well as absorbers. A lower 
velocity limit of $cz \geq 2500$ km~s$^{-1}$ was required 
so that the \lya \ line could 
easily be differentiated from the bright geocoronal \lya\ line in any 
spectroscopic observations.
The sample was observed during 30 orbits in Cycle 11. The observations were
obtained with the F122M filter for \lya \ online imaging and F140LP as a
continuum filter.

The first results for two of the observed galaxies, ESO\,350-IG38 and 
SBS\,0335-052, were presented in Kunth et al. (2003). 
In that study, we subtracted the continuum by
using the relative instrument 
sensitivities (PHOTFLAM values) of the two instrument configurations, 
using the continuum slope ($\beta$) obtained from IUE or GHRS data 
for the bright knots and assuming a flat continuum in all other regions. 
However, these assumptions are not sufficient to perform an accurate
photometric analysis and allowed the authors only to obtain a
qualitative picture of the \lya \ morphology.

In this paper we focus on the most nearby galaxy in the sample, 
the well known luminous blue compact galaxy ESO\,338-IG04.
For this galaxy, Wide Field and Planetary Camera 2 (WFPC2) images in numerous
optical passbands and Space Telescope Imaging Spectrograph (STIS)
spectroscopy are also available. Using
these we develop methods for accurate continuum subtraction, allowing a
quantitative analysis of the \lya \ emission and absorption. 

 \subsection{Short facts about the target: ESO\,338-IG04}

\begin{figure}
\resizebox{\hsize}{!}{\includegraphics{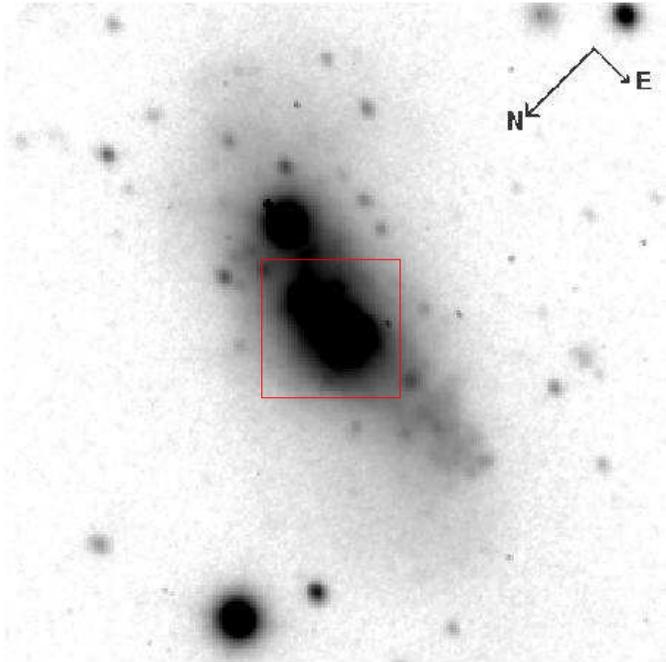}}
\caption{Ground based R-band image of ESO\,338-IG04 taken with the EMMI 
instrument of the NTT on La Silla (Bergvall \& \"Ostlin). 
The image is 70 $\times$ 70\arcsec\ with the longer north
arrow of the compass representing 10\arcsec. The image has been rotated to
the same orientation as the ACS images. The box in the centre of the galaxy
shows the 15 $\times$ 15\arcsec\ region shown in figure \ref{specim}.} 
\label{figgroundbased} 
\end{figure}

ESO\,338-IG04 (Tololo\,1924-416) is a well known luminous $(M_V=-19.3)$ 
blue compact galaxy (BCG), in the local universe 
($v_{\rm rad}=2830$ km~s$^{-1}$, Cannon et al, 2004) and 
can be seen in figure \ref{figgroundbased}.
The galaxy was first studied in detail by Bergvall (1985). The nebular oxygen
abundance is $12+\log({\rm O/H})\approx 8.0$ (see \"Ostlin et al. 2003).

Images obtained with HST's Faint Object Camera (FOC) show that the centre of 
ESO\,338-IG04 hosts a population of UV-bright star clusters, just like many 
other starbursts (Meurer et al. 1995). Subsequent 
WFPC2 observations have increased the number of known clusters to more than 
100 and have shown them to span ages from just a few Myr to several Gyr 
(\"Ostlin et al. 1998, 2003). The age distribution of the young clusters
indicates that the present starburst was initiated 40 Myr ago. 
Spectroscopic observations from the International Ultraviolet Explorer (IUE) 
showed ESO\,338-IG04 to be a bright \lya \ emitter with a UV continuum that 
has close to zero internal dust extinction (Giavalisco et al. 1996, 
Calzetti et al. 1994).

ESO\,338-IG04 has a perturbed H$\alpha$ velocity field (\"Ostlin et al. 1999, 
2001) and morphology (Bergvall and \"Ostlin 2002) suggesting that the galaxy 
has been involved in a merger/strong interaction with a small galaxy. The 
presence of a companion galaxy (\"Ostlin  et al. 1999) and extended H{\sc i} 
emission between the two (Cannon et al. 2004) further supports this starburst
triggering scenario.
	The galaxy is a true starburst in the sense that the time-scales for
gas consumption and build-up of the observed stellar mass with the current
SFR are both $\sim 1$ Gyr, i.e. much shorter than the age of the universe
(\"Ostlin et al. 2001).

The bright \lya \ line, the existence of HST data at other wavelengths and the
small internal reddening made ESO\,338-IG04 a perfect target for a \lya \ 
imaging study with ACS.

\section{Data acquisition and basic reductions}

\subsection{ACS observations}

The observations were obtained with the Solar Blind Channel (SBC) of the
ACS under general observer (GO) program 9470, using two filters: 
F122M (\lya \ on-line) and F140LP (used as continuum filter). The ACS/SBC was
the ideal instrument mode for these observations because its field of view is 
appropriate to encompass the galaxies comfortably and utilises a MAMA detector
and is hence not susceptible to cosmic rays.
The target galaxy was observed for 5 orbits, with the F122M images 
being obtained during the SHADOW part of the orbit in order to lower the 
geocoronal background. The remainder of the orbit was then used for the F140LP
images. The geocoronal background is low in both filters but is 
almost always brighter than the signal in the online
image from the target galaxies.
The total integration times per galaxy in the F122M and F140LP filters 
were 9095 and 3000 
seconds respectively. Observations in each filter were split into five
separate exposures.
The drizzled images produced by the standard pipeline have a rectified
pixel scale of 0.025\arcsec~pix$^{-1}$. 
These images were combined after coalignment, 
and the background in each filter was subtracted using a first degree
polynomial surface. These images were then corrected for Galactic interstellar
extinction using the Cardelli law (Cardelli et al. 1989). Values used were
${E(B-V)} = 0.087$ and $A_{B}$=0.375 (Schlegel et al, 1998). 
The solid angle subtended 
by ESO\,338-IG04 is so small that these values can be assumed over the 
entirety of the images.


\subsection{Supporting HST observations}

In addition to the ACS images we make use of images in the  F218W, F336W, 
F439W, F555W and F814W filters obtained with the Planetary Camera 
(PC) aperture of the WFPC2 under GO 
program 6708. These observations have been used to study bright 
compact star clusters and are described in \"Ostlin et al (1998, 
2003). We use this images aligned and rebinned to the same orientation
and scale as the ACS data. These images were also corrected for Galactic 
interstellar extinction at the central wavelengths of the filters. 


A narrowband image of the 
[O{\sc iii}]$_{{\rm \lambda}4959\AA}$ emission line has 
been obtained with HST/STIS in GO program 8668
(\"Ostlin et al. in prep). This observation takes advantage of the 
fortuitous redshift of ESO\,338-IG04 where the 4959\AA\ line is redshifted
to 5007\AA\ and can be imaged with the very narrow (FWHM = 5\AA) 
STIS/F28X50O{\sc iii} filter centred on 5007\AA.

\begin{figure}
\resizebox{\hsize}{!}{\includegraphics{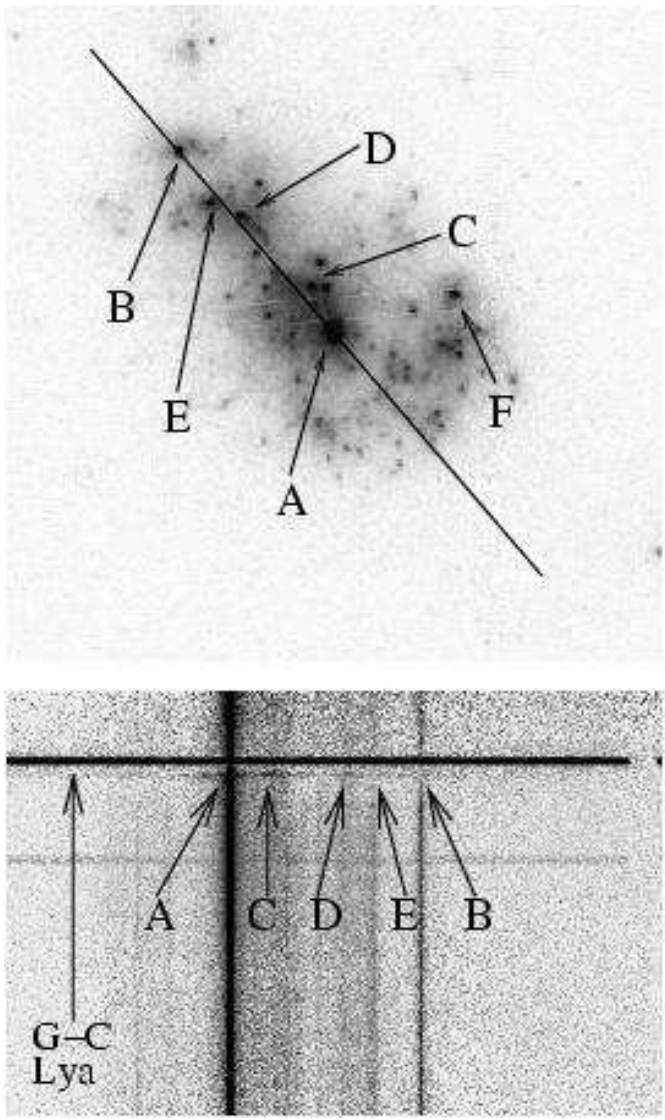}}   
\caption{
{\bf Top}: ACS/F140LP image with the STIS slit position overlayed
and certain features labeled. The image size is $15 \times 15$\arcsec. 
This image corresponds in size and orientation to the box in figure  
\ref{figgroundbased} and shows how the bright UV continuum radiation 
emanates from a much smaller region than the observed size of ESO\,338-IG04
in optical bandpasses.
The nearly horizontal lines running across the centre of the image are 
due to the mosaic of the detector and couldn't quite be removed with our 
dithering pattern. 
{\bf Bottom}: STIS G140L spectral image for the
slit position indicated above. $\lambda$ decreases with increasing 
ordinate. Two geocoronal lines are visible running horizontally: 
the strong \lya\ line can be seen towards the top of the image and the 
fainter O{\sc i} line at 1304\AA\ further down. 
The labels denote the same spatial regions as in the upper panel
 and point at 
the \lya \ wavelength in the restframe of the target. 
} 
\label{specim} 
\end{figure}

We also make use of a HST long-slit spectrum obtained with the
Space Telescope Imaging Spectrograph (STIS) with the G140L grating
and a 0.2\arcsec \ wide slit, under GO program 9036. The slit was 
placed approximately along the major axis of the target as shown in 
figure \ref{specim}.

\section{Subtracting the continuum}

We use the F140LP filter to subtract the continuum from F122M and produce
a line-only \lya \ image. 
As with any continuum subtraction, the continuum
image is used to estimate the flux due to continuum processes at the 
wavelength of the line. It is necessary to scale the continuum image to
compute the continuum level at the the line. 
We define the Continuum Throughput Normalisation (CTN) factor as the ratio of 
flux in the F140LP to that in the  F122M filter for a given spectral 
energy distribution (SED). 
Hence with no \lya \ feature in emission or absorption: 
\begin{center}
\[  (f_{F122M} \times {\rm CTN}) - f_{F140LP} = 0  \]
\end{center}
where $f_{F122M}$ and $f_{F140LP}$ are the $f_{\lambda}$ 
fluxes in the F122M and F140LP  filters respectively.

If one knows the instrument sensitivities in these filters and the
SED between these wavelengths, the CTN 
factor can readily be computed.
The simplest method to obtain a reasonable first order estimate of CTN factor
 is to assume the SED
to be a power-law governed by the index \bet \ such that 
$f_{\lambda}=\lambda^{\beta}$.
Kunth et al. (2004) used such an assumption to make \lya\ continuum 
subtractions for the first two galaxies in the sample,  ESO\,350-IG38 
and SBS\,0335-053. There we used available IUE spectra and WFPC2 V-band images
to estimate global values of \bet\ and used these values to 
subtract the continuum in the regions dominated by bright clusters.
Assuming standard calibration and a flat 
continuum $(\beta=0)$, the CTN is simply the ratio of the 
PHOTFLAM keywords of the off- and 
online instrument configurations: 10.33. This value was then used to subtract
the continuum in the remaining (non star-cluster) regions. 
As a first approximation, this method provides a reasonable map of 
\lya\ emission and absorption and enabled the authors to discuss the
kinematics of the line. It is, however, highly dependent upon the validity
of the assumptions and doesn't allow any accurate photometric study to be 
performed. Furthermore this method doesn't properly allow for spatial
variation in the 
slope of the continuum -- a quantity that is highly sensitive to local 
variations in burst age and internal dust reddening. 

\begin{figure}
\resizebox{\hsize}{!}{\rotatebox{-90}{\includegraphics{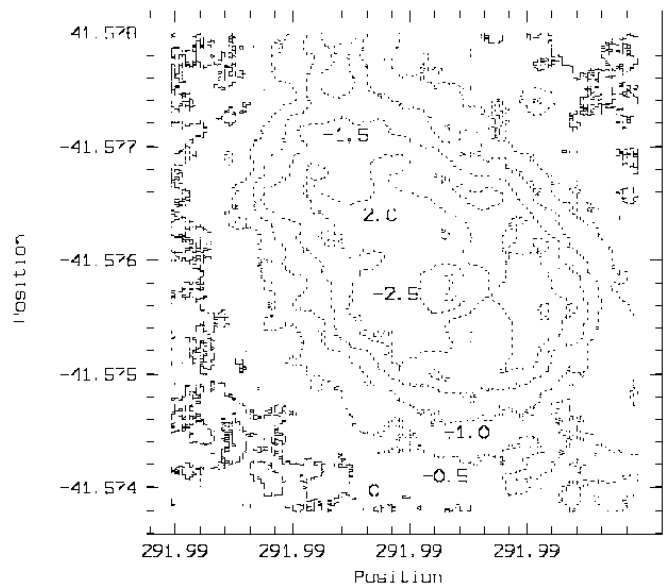}}}
\caption{Contour plot showing the spatial variation of the UV continuum
slope \bet \ between the ACS/F140LP filter at 1527\AA \ and the 
WFPC2/F218W filter at 2204\AA. 
The innermost contour corresponds to the steepest slope of
\bet=-3. Subsequent contours are spaced at equal intervals of 0.5. 
The image covers $15 \times 15$ \arcsec. The image has been smoothed using 
a $10 \times 10$ pixel boxcar filter.} 
\label{beta}
\end{figure}

For ESO\,338-IG04, the UV continuum slope was mapped by using the ACS/F140LP
image and a variety of WFPC2 images at longer wavelength.
A resulting \bet-map, constructed using the F218W filter can be seen in figure 
\ref{beta}. As the map shows, \bet \ reaches a minimum of -3 in the 
central bright star cluster, indicating a steep continuum dominated by 
O and B stars. \bet \  increases to values of 
$-0.5 \lesssim \beta \lesssim 0.5$ away from the starburst region, i.e. an 
approximately flat continuum. 

Using such a map of the continuum slope to subtract the continuum assumes
that the UV continuum between 1527\AA \ and \lya \ remains an unbroken power
law. Preliminary continuum subtractions using this technique showed strong
\lya\ absorption throughout the regions of bright continuum. This is 
contradictory to the STIS spectra extracted from these regions which show
\lya \  in emission and 
is indicative of the fact that this technique causes the subtraction of too
much continuum. This discrepancy can be attributed to a number of effects:
Firstly, examination of the IUE spectrum suggests that the continuum slope 
deviates from the power-law, flattening out at wavelengths shortward 
of $\sim 1400$\AA; secondly, the presence of discrete stellar \lya\ absorption
features may also serve to reduce the total \lya \ output from a region. 
Often most importantly, inspection of the STIS spectra shows that Galactic 
absorption of \lya\ is significant in the regions where the continuum is
strong, causing a significant reduction of the flux in the online filter.
While the assumption of a power-law continuum may be sound redwards of 
\lya, it fails to address any effects that may arise as a result of 
Galactic H{\sc i} and some technique that accounts for this is essential.

\subsection{Synthetic spectra}

In order to tackle some of the issues raised above, we began to explore 
the properties of the Starburst99 synthetic spectral models (Leitherer
er al, 1999). 
These models were selected because they use the most appropriate UV stellar 
atmospheres and  cover a large parameter space enabling us
to test how different model parameters
may affect our final results. The possibility also exists to run one's own 
simulations, based upon exact requirements and to interface with the 
Mappings code (Kewley et al. in prep) for the inclusion of nebular
emission lines. The initial model set selected included stellar and nebular
emission for an instantaneous burst of metallicity $Z=0.001$ and 
Salpeter IMF. The metallicity of ESO\,338-IG04 is known to be around 10\% solar
(Bergvall 1984; Masegosa et al, 1994) so the lowest available metallicity 
was chosen. \"Ostlin et al (2003) showed that a Salpeter IMF extending up 
to $\sim 100M_{\odot}$ was the best fit to their data on this galaxy. 
The effects of varying these parameters are addressed in section 5. 
The models were used to investigate the CTN factor and optical 
colours with the effects of aging and dust reddening.
 
\begin{figure}
\resizebox{\hsize}{!}{\includegraphics[angle=0]{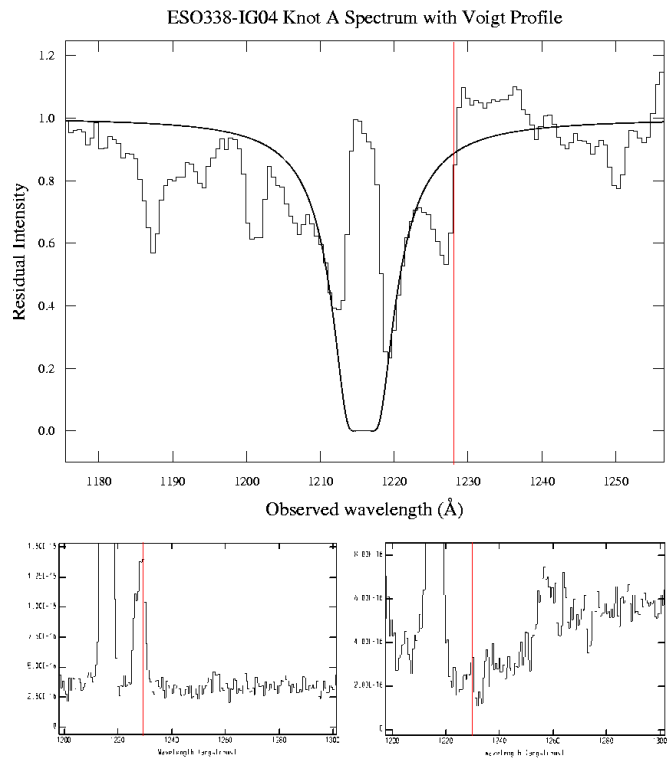}}
\caption{
{\bf Top}: Part of the STIS G140L spectrum for knot A  with a Voigt
profile fitted to the wings of the damped Galactic \lya\ absorption 
profile, giving a Galactic HI column density of 20.7 cm$^{-2}$. The strong
geocoronal \lya \ line can clearly be seen at 1216 \AA. The vertical line
at 1228 \AA\ shows the wavelength of \lya\ in the target galaxy and 
blueshifted absorption can clearly be seen just shortwards of this wavelength.
Flux units are normalised to the continuum at 1300\AA. 
{\bf Bottom}: STIS spectra covering 1200 to 1300\AA\ 
from knots C (left) and B (right) where \lya\ is seen in emission
and absorption respectively. 
The absorption trough extends from around the geocoronal line to 
wavelengths as long as 1250\AA.
} 
\label{voigt} 
\end{figure}

One feature that should be the same in all spectra is the Galactic \lya\
absorption which removes a fixed fraction of photons near 1216\AA. 
Since ESO\,338-IG04 covers little solid angle on the sky (less than half a square 
arcmin), we assume the column density of Galactic H{\sc i} along the 
sightline is constant over the face of the galaxy. A 1D STIS spectrum of the 
central star cluster (where signal-to-noise is highest) was extracted.
Using the {\sc Xvoigt} program (Mar \& Bailey, 1995) we made a single-component
fit of a Voigt profile to the wings of the Galactic absorption
profile. This yielded a Galactic H{\sc i} column density of 
log(N(H{\sc i}) = 20.7 ${\rm cm}^{-2}$, in tight agreement
with the value of log(N(H{\sc i})) = 20.8 ${\rm cm}^{-2}$ 
from Galactic H{\sc i} maps of Dickey \& Lockman (1990). This normalised
Voigt profile was then used to apply the Galactic \lya\  
absorption profile to the
synthetic spectra by simple convolution. 
A fit of the Voigt profile to the wings of the 
absorption profile can be seen in the upper panel of figure \ref{voigt}.

Galactic interstellar reddening also has the same affect on any spectrum since
reddening simply removes a fraction of the flux at a given wavelength. 
The Galactic extinction law of Cardelli et al (1989) was used with the 
parameters of $A_{\rm V} = 0.288$ and $E(B - V) = 0.087$ (taken from the 
values quoted in NED, calculated from Schlegel, 1998) to redden all the spectra. 

Our investigation was mainly concerned with how age and
internal reddening affect the CTN factor and 
broadband UV/optical colours when computed from synthetic spectra.
We obtained the instrument throughput sensitivity profiles for all
of our instrument configurations from the {\sc synphot} package in the
{\sc iraf/stsdas}. This allowed us to compute synthetic fluxes from the 
spectra in all of our passbands by simple integration of the spectrum
convolved with the instrument sensitivity profile. Conversion to the 
Vega magnitude system was performed using the quoted zeropoint offsets. 
Since emission lines are not included in the spectra (no \lya ),
the CTN factor can readily be computed as the ratio of the 
synthetic flux in the offline filter to that in \lya \ filter (see 
the equation).

While the calculation of the CTN factor from a spectrum may be 
computationally trivial, knowing exactly which spectrum to use 
(i.e. which age and internal reddening) is far more complex. 
The region of interest is a bright, young region of intense
star formation where the age of young clusters ranges from 1 to $\sim40$Myr 
(\"Ostlin et al, 2003). 
Studies of the central starburst region using the Balmer emission
line decrement (eg Bergvall 1985, Raimann et al. 2000) and the UV 
continuum (Calzetti et al. 1994, Meurer et al. 1995, Buat et al 2002) show
the internal reddening to be very low: $E(B-V)<0.05$. Furthermore, 
groundbased longslit spectroscopy measuring H$\alpha$/H$\beta$ 
(\"Ostlin et al, 2003) determined $E(B-V) \le 0.20$ everywhere 
along the slit (aligned in the east-west direction).

Due to the well-known degeneracy in the  effects of age and reddening, 
it is not
possible to derive a non-degenerate relationship between a single 
colour (eg. $U-B$ or $\beta$) and the CTN factor. 
Instead we began to look for a possible 
relationship between CTN factor and {\em pairs} of colours. 
Our goal was to establish a non-degenerate
relationship between CTN factor and any pair of colours obtained from the
SED, 
irrespective of the age of the starburst and the internal dust reddening. 
The procedure we implemented was as follows:
\begin{enumerate}
\item{
Begin with the full age range (1 to 900 Myr) of Starburst99 spectra with
Galactic reddening and H{\sc i} absorption profiles applied. 
}
\item{
Redden each spectrum using the SMC law (Pr\'evot et al., 1984). Values of 
internal colour excess were in the range $E(B - V)$ 0 to 0.4 in steps of 
0.003 mag; extending beyond the maximum observed reddening of 0.2 mag. 
}
\item{
For each spectrum (many ages, many internal reddenings), convolve with 
all instrument throughput profiles and integrate to obtain synthetic 
fluxes. From these, compute CTN factors and colours (convert
to Vega magnitude system).
}
\item{
Linearly interpolate extra points for CTN factor and colour in
between the jumps in age. 
}

\end{enumerate}

\begin{figure*}
\resizebox{\hsize}{!}{\includegraphics{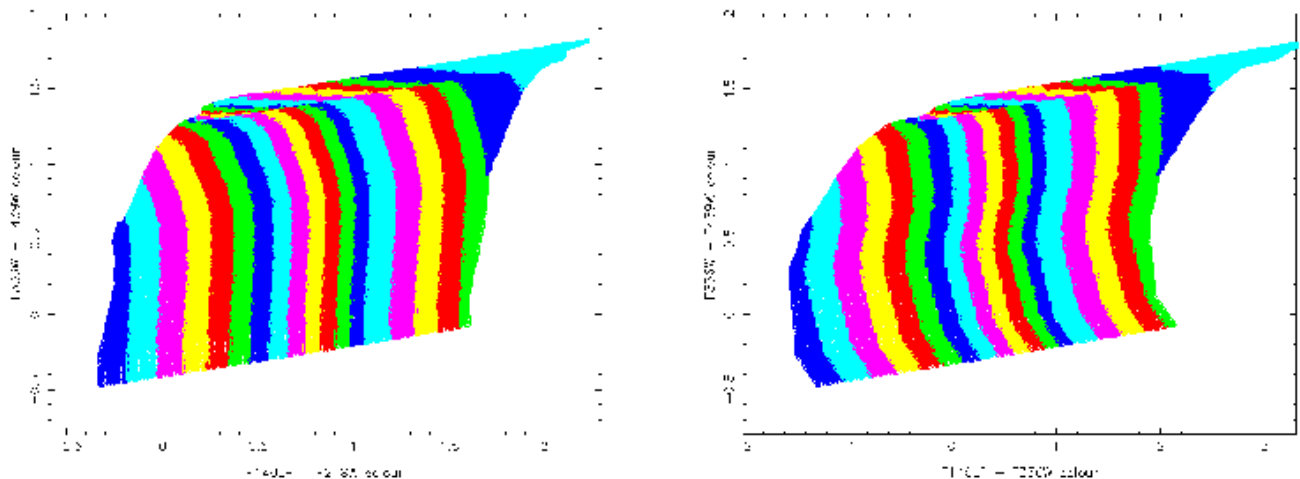}}
\caption{
Plots using two synthetic colours from the Starburst99 models. 
The CTN factor is represented by
colour in bins of width 1 unit. 
7-8 is shown in blue in the bottom left (bluest corner),
8-9 in the neighbouring cyan bin, 
9-10 in magenta, etc...
{\bf Left}: F140LP--V218W vs. F336W--F439W. 
{\bf Right}: F140LP--F336W vs F336W--F439W. 
Colours are in Vega magnitudes.} 
\label{colcolx} 
\end{figure*}

This procedure resulted in a large table of values: a CTN factor and 
all UV/optical colours for every step in $E(B-V)$ for every age. 
To look for relationships between the colours and CTN factors, we produced
a series of colour-colour plots. For each possible age and reddening, a point 
was plotted in colour-colour space where the CTN factor was represented by 
the colour of the point. One plot was produced for every possible combination
of HST colours. Each plot was studied to look for a combination of colours 
that corresponded to a single, non-degenerate value of the CTN factor.
Two such diagrams are shown in figure \ref{colcolx}.

It can be seen in figure \ref{colcolx} that for these combinations of 
colours, if the ordinate and abscissa are known, they can be used to find
a single, non-degenerate value of the CTN factor. The F140LP-F336W vs.
F336W-F439W combination was then chosen to produce a CTN factor map. 
While producing nice-looking plots, any set-ups involving the WFPC2/F218W
filter were not chosen due to poor S/N in this image. 
In areas of low S/N, the chosen images were smoothed using a $10\times10$
pixel box average filter. 
For each pixel in the aligned images,
F140LP-F336W and F336W-F439W colours were calculated.
These colours were converted to a CTN
factor and written back to the corresponding pixel of a new image.
This CTN factor map was
then used to create the desired \lya\ line-only image.

\section{Results}
 
\begin{figure*}
\resizebox{0.98\hsize}{!}{\includegraphics{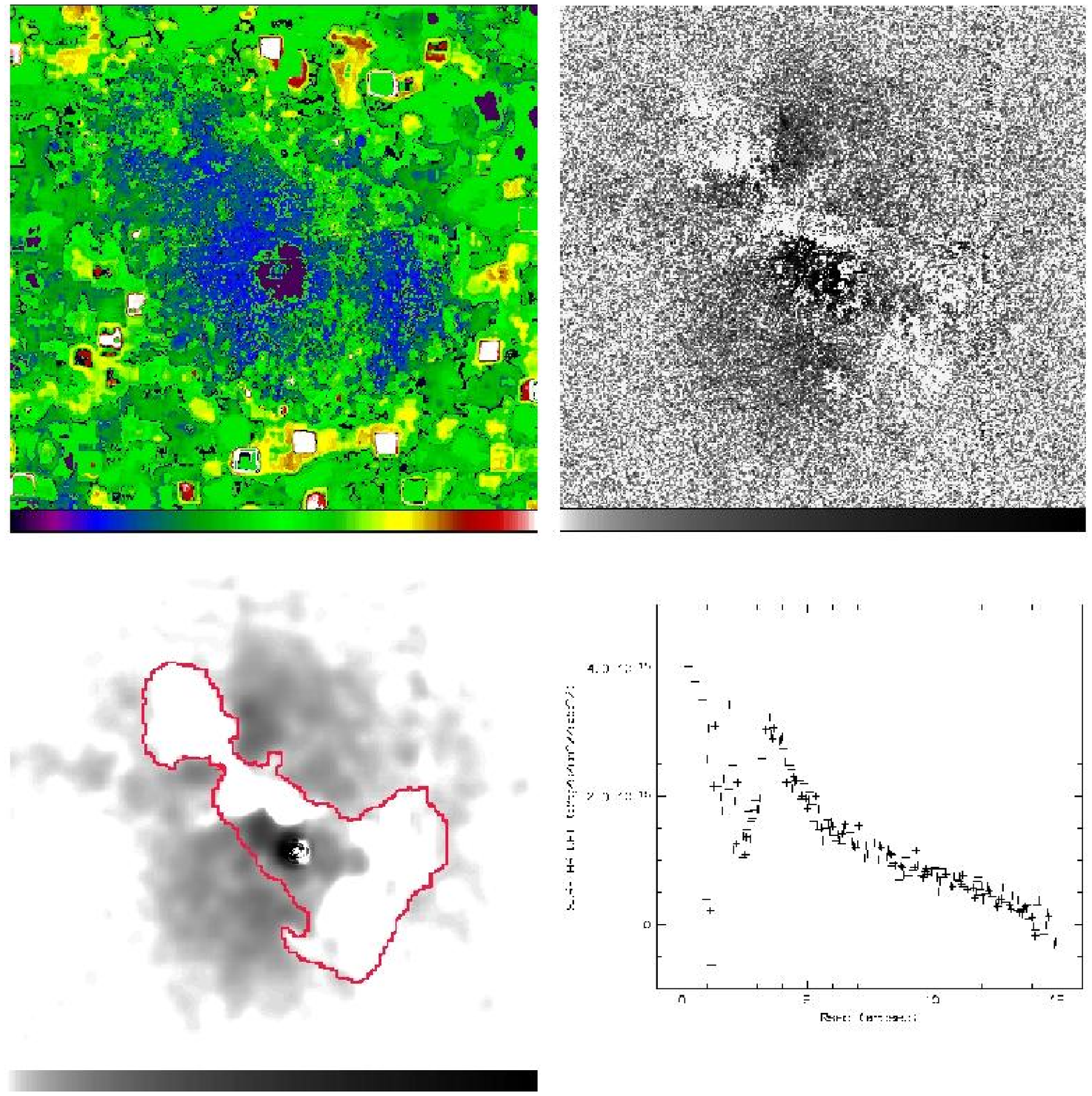}}
\caption{{\bf Top left}: Map of CTN factor 
across the starburst region of ESO338--IG04. 
The intensity scale ranges from 7 in blue/purple
to 14 in red/white. The midpoint green colour surrounding the starburst
therefore corresponds to CTN $\simeq 10.5$. 
A flat $(\beta=0)$ power-law continuum corresponds to CTN$=10.33$. 
This image has been smoothed in
the outskirts where S/N is low using a $10 \times 10$ pixel boxcar 
filter.
{\bf Top right}: \lya \ image using the CTN map to the left for the continuum 
subtraction. Inverted log scale so emission is black, absorption white. 
\lya\ escapes from the centre due to outflowing gas
and is observed in diffuse regions where photons escape after  H{\sc i}
scatterings.
{\bf Bottom left}: Map of the \lya\ absorption and emission, smoothed using
an adaptive filtering algorithm. The red line indicates the border of the 
mask used to examine the central and diffuse regions individually.
 The smoothed version clearly shows knots A
and C as spatially separate \lya\ sources.
The size of the images is $15 \times 15$ \arcsec.
{\bf Bottom right}: Radial distribution of \lya \ luminosity in ESO\,338-IG04. 
Points in the innermost 0.5\arcsec\ fall above the plot. The profile shows 
a peak in \lya\ surface brightness at 3.5\arcsec from the centre, 
corresponding to the diffuse emission regions seen in the top right and bottom
left panels.} 
\label{results} 
\end{figure*}

The first CTN-factor map and the continuum subtracted \lya \ image
can be seen in figure \ref{results}. The green regions surrounding 
the starburst show the CTN factor to fall in the range 10.0 to 10.5. This 
is consistent with the flat continuum slope which can also be seen in figure
\ref{beta} where  \bet\ is shown to fall in the range -0.5 to 0.0.
In the centre the CTN factor falls to 
the minimum value permitted by the models: 7.2 (1Myr, unreddened burst).

The first continuum subtracted \lya\ image, in the right panel of fig
\ref{results} shows excellent qualitative agreement with features in
the spectrum. Most notably: 

\begin{itemize}
\item \lya\ is seen in both emission and absorption along the slit.
See also figures \ref{specim} and \ref{voigt}.
\item Evidence for both emission and absorption can be seen in
the regions immediately surrounding knot A. 
Bright emission and a small absorption hole appear in the continuum 
subtracted image. This is consistent with the P-Cygni profile, centred
around 1228\AA\ visible in figure \ref{voigt}.
\item Emission is seen from the regions around knots C and D 
(consistent with the STIS spectrum). The lack of emission from these regions
was one of the serious inadequacies of the \bet -map technique.
\item Absorption is seen around knots B and E. Furthermore, the CTN
factor map shows a decrease in CTN factor in the region of these knots.
This is consistent with the continuum flux decreasing towards \lya \ in
these knots as can be seen in the spectral image (figures \ref{specim} \& \ref{voigt}).
\item Although this is not covered by the STIS spectrum, diffuse emission 
is detected from large regions outside the central starburst. 
\end{itemize}

Quantitative photometry was performed on the continuum subtracted \lya \ 
image and a total \lya\ flux 
$f_{\rm Ly\alpha,TOT} = 194 \times 10^{-14}$erg~s$^{-1}$~cm$^{-2}$ was
obtained. This is the sum of all the flux within a radius of 15\arcsec, 
at which distance the \lya\ surface brightness falls to zero.
\lya\ images were then masked with a 10 $\times$ 20 \arcsec\ elliptical mask, 
centred on the central star cluster (knot A). This corresponds to 
the size and shape of the aperture of the IUE satellite, from which 
the integrated \lya\ flux and equivalent width of ESO\,338-IG04 was presented
in Giavalisco et al. (1996).
These authors claim a \lya \ flux of $123\times 10^{-14}$ 
erg~s$^{-1}$~cm$^{-2}$. 
Analysis of our continuum subtracted \lya \ image reveals a \lya \ 
flux of 
$f_{\rm Ly\alpha,IUE} = 134\times 10^{-14}$ erg~s$^{-1}$~cm$^{-2}$, perfectly 
consistent, considering the accuracy of the IUE. 
At a distance of 37Mpc, this corresponds to \lya\ luminosity 
L$_{{\rm Ly} \alpha}= 3.17 \times 10^{41}$~erg~s$^{-1}$.

The surface brightness profile of the \lya\ emission line can be seen in 
figure \ref{results}. This profile was created by integrating the flux
outwards from the central star cluster in concentric circles.

\begin{figure*}
\resizebox{0.98\hsize}{!}{\includegraphics{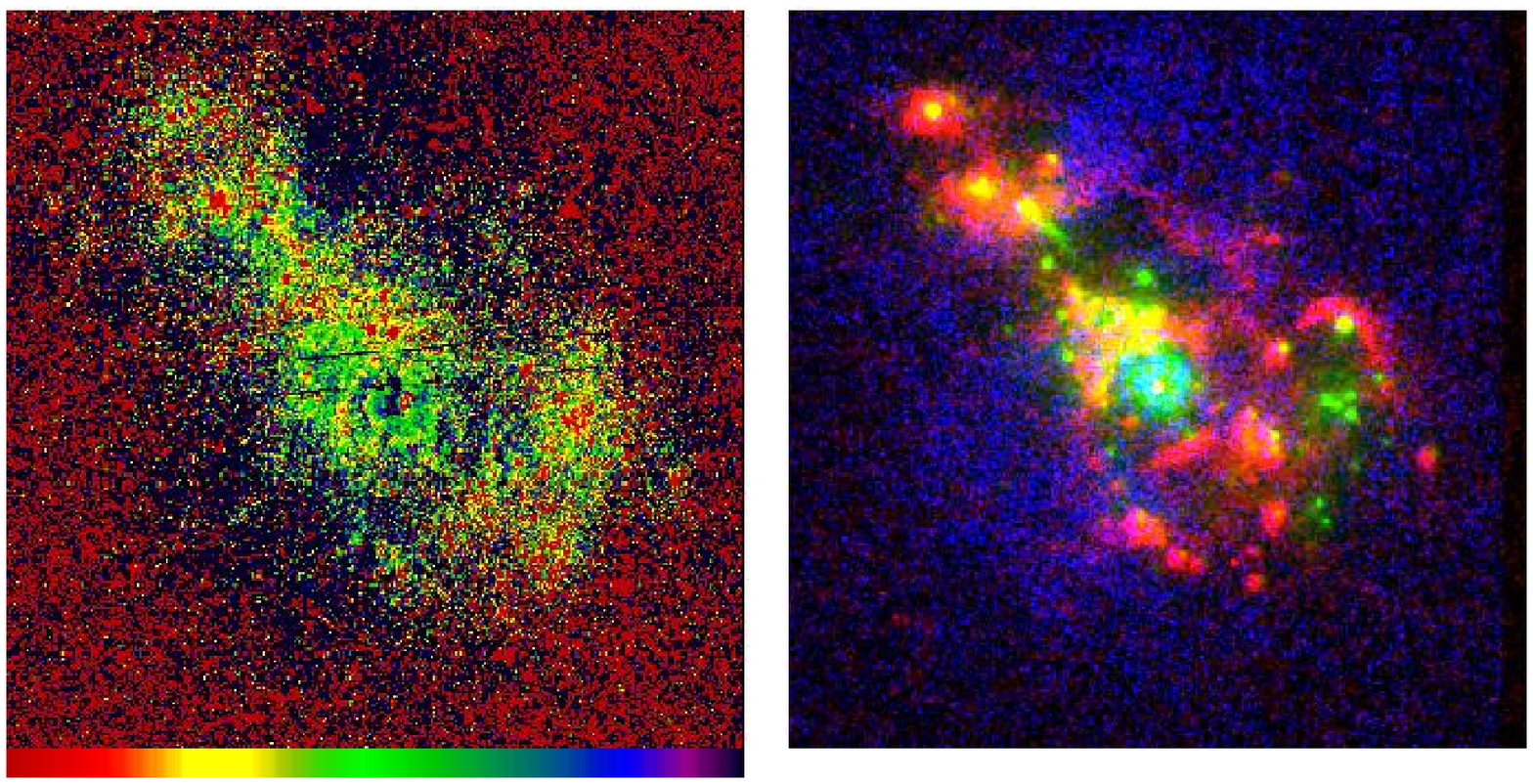}}
\caption{
{\bf Left}: Equivalent width map -- negative scale -- cuts levels are 
-50 to +50. Regions of high equivalent width show up in dark colours.
Particularly visible are the diffuse emission regions outside the starburst
region. Much local structure can be seen, particularly around knot A and 
the other bright continuum sources.
{\bf Right}: A false colour image: red shows the [O{\sc iii}] STIS image; 
green shows the UV continuum F140LP image; blue the continuum subtracted
\lya \ image. Cuts levels are set to boost interesting features.} 
\label{colew}
\end{figure*}

Using the CTN map, the F140LP image was renormalised to the continuum level at
\lya, and used together with the online image to create an equivalent width
map. This map can be seen in the left panel of figure \ref{colew}.
Using the same IUE aperture mask as before, the equivalent width
of \lya \ in this aperture was calculated as 22.6\AA. This again corresponds
very closely to the value of 26.2\AA\ presented in Giavalisco et al 
(1996). The equivalent width  exceeds 200\AA\  in the diffuse
emission regions where there is little continuum. 

The equivalent widths were investigated in detail in some of the absorption 
regions. Absorption equivalent widths in some regions around the brightest
star clusters are as low as -65\AA, indicating that the H{\sc i} absorption
absorption in these regions is damped. 

The RGB composite shown in figure \ref{colew} is composed of the continuum
subtracted \lya \ image created with the assumed parameters, the F140LP ACS 
image continuum image and the [O{\sc iii}] STIS image. Colour channels are: 
Red -- [O{\sc iii}] ; Green -- UV ; Blue -- \lya. This is discussed in section 6.

\section{Model dependencies and other considerations}

The results presented above utilised a our assumed set of model spectra, 
based upon various parameters relating to the  properties of
the target galaxy. 
The Starburst99 spectra we used 
were created for very low metallicity and Salpeter IMF, with the additional
assumption of the SMC law being made to redden the spectra in the 
procedure of mapping the CTN factor. 
Some parameters were chosen based upon 
various known facts about the target galaxy (eg the low metallicity and 
the IMF) and others were assumed based upon studies of similar galaxies
(eg the SMC law for internal reddening).
If we want to know to what extent our results can be trusted, it is important
to know how the selection of individual model parameters may influence our 
studies. 

Some parameters are easily tested by changing the set of model spectra. We 
used  other  Starburst99 synthetic spectra to investigate the effects 
of metallicity and IMF, simply by plugging a different set of models into
our code. 
Similarly, changing the assumed law for internal reddening law was trivial
to implement by making minor modifications to the software. We adopted the 
assumed set of model parameters ($Z=0.001$, Salpeter IMF, SMC law) as 
the ``standard'' set, to which we compared all our 
subsequent modifications. 
By changing one parameter at a time, we recomputed CTN maps and used them
to create continuum subtractions. We compared the individual pixel values of the 
modified CTN maps to the pixels of the standard CTN map and performed 
the same quantitative photometry on the \lya\ line-only images as described 
in section 4. Masks isolating the central and diffuse regions were also created
so that these components could be examined individually if necessary. 
The parameter space explored in this manner is listed in table
\ref{tabmodelparsall}.

\begin{table}[h]
\caption[]{Model parameters for consideration} 
\begin{flushleft}
\begin{tabular}{cccc} 
\hline
\noalign{\smallskip}

Red. law & Metallicity  & IMF $\alpha$     & IMF $M_{\rm up}$           \cr
\hline
\hline
SMC      &  0.001 & -2.35 & 100 \cr
\hline
LMC      &  0.001 & -2.35 & 100 \cr
Cardelli &  0.001 & -2.35 & 100 \cr
Calzetti &  0.001 & -2.35 & 100 \cr
\hline
SMC      &  0.004 & -2.35 & 100 \cr 
SMC      &  0.008 & -2.35 & 100 \cr 
SMC      &  0.020 & -2.35 & 100 \cr
SMC      &  0.040 & -2.35 & 100 \cr
\hline
SMC      &  0.001 & -3.3  & 100 \cr
SMC      &  0.001 & -1.5  & 100 \cr
SMC      &  0.001 & -2.35 & 30  \cr
\hline
\end{tabular}\\
Notes: 
$\alpha$: $dN \propto M^{-\alpha} dM$, 
$M_{\rm low}=1M_{\odot}$.
\end{flushleft}
\label{tabmodelparsall}
\end{table}

Other effects that were investigated include the effects of 
nebular emission lines, a different spatial distribution of the
nebular emission with respect to the ionising continuum, continuous 
star formation, and the stellar \lya\ absorption feature. 

It is important to note that, while changing model parameters may have a 
quantitative effect on the \lya\ flux and equivalent width, no noticeable 
differences could be seen by eye in the continuum subtracted \lya\ images.
Morphologically, the images resemble one another.


\subsection{Metallicity and IMF parameters}

\begin{figure}
\resizebox{0.85\hsize}{!}{\rotatebox{0}{\includegraphics{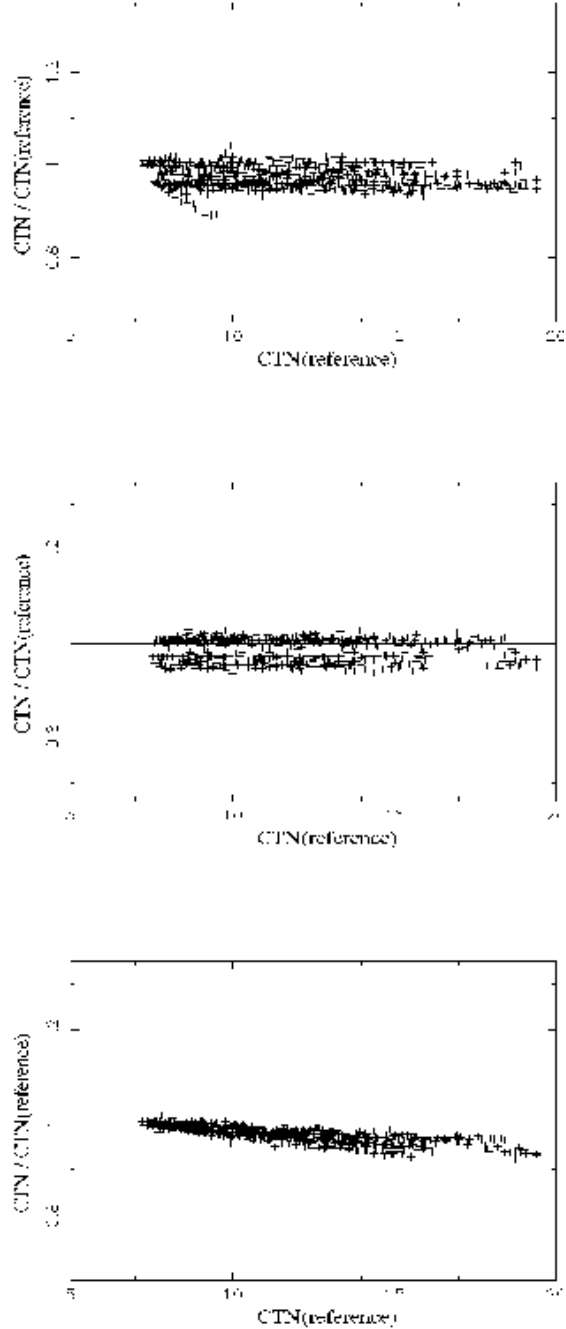}}} 
\caption{Figures showing the effects of varying certain model parameters. 
Each point represents one pixel in the CTN map. The abscissa represents 
the value computed with the assumed set of model parameters. The ordinate 
shows the value of that pixel, normalised by the abscissa. I.e. the ordinate
is unity if the pixel is unchanged by changing model parameters. The perturbed 
parameters are: 
{\bf Top}: Metallicity $(Z=0.008)$;
{\bf Centre}: IMF $(\alpha = -3.3)$;
{\bf Bottom}: Internal reddening law (LMC).}
\label{varypara} 
\end{figure}

Table \ref{tabmodelparsall} shows metallicity and IMFs investigated. 
Examples
of how these parameters affect individual pixels in the CTN map
can be seen in the top two panels of figure \ref{varypara}. 
The value of each pixel in 
the ``standard'' map is represented by the abscissa of each point with 
the value of the pixel in the new CNT map, normalised by its standard value,
is represented by the ordinate. 

It can be seen from the top plot that increasing the metallicity from 
$Z=0.001$ to $Z=0.008$ (well above the measured value) 
introduces a slight downward trend in the CTN
values. This effect was present for all metallicities shown in table 
\ref{tabmodelparsall}. Increasing the metallicity to extreme values 
of $Z=0.040$ produced similar plots but with the scatter in the 
ordinate increased. Selection of a metallicity of $Z=0.008$ resulted 
in a decrease in the \lya\ flux to 
$f_{\rm Ly\alpha,IUE} = 113\times 10^{-14}$ erg~s$^{-1}$~cm$^{-2}$, 
a drop of 16\%.

The centre plot in figure \ref{varypara} shows the effect of steepening the
index of the power-law governing the IMF to 
$\alpha = -3.3$. The effect is very similar to that obtained by 
increasing the metallicity: a slight, systematic decrease of CTN. The same
result was obtained when using the model set with 
$\alpha = -2.35, M_{\rm up} = 30 M_{\odot}$.
Both modifications to the IMF have similar effects on the CTN due to the fact
that such modifications have similar effects on the massive stellar
population: the relative fraction of massive stars is reduced, flattening 
the continuum in the UV. The use of a flatter IMF with $\alpha = -1.5$ had
the opposite effect: the \lya\ flux increased over that obtained 
with our standard parameter set to 
168 $\times 10^{-14}$~erg~s$^{-1}$~cm$^{-2}$.

It can be seen that increasing metallicity and decreasing the relative 
number of hot young stars in the IMF have similar effects on the individual
pixels in the CTN maps: a general, systematic decrease. 
This is an interesting result
because one could naively assert that a process that flattens the continuum
should lead to a {\em increase} in the CTN since the F140LP / F122M ratio will
increase. However, as well as the CTN
itself, metallicity and IMF also affect the  optical/UV colours --
particularly the F140LP-F336W colour which represents the slope of the UV continuum.
A redder SED leads to an 
increased CTN but also to a higher value for the F140LP-F336W colour
which serves to shift the points in the right panel of figure \ref{colcolx} 
to the right. Hence the colour obtained from the optical images may actually 
correspond to a lower CTN when the spectrum becomes flatter. It all depends
upon whether the effect under consideration has a stronger impact on the 
CTN factor or the optical colours.

\subsection{Extinction law}

In the method described in section 3, we assumed the SMC law of
Pr\'evot et al. (1984) 
\cite{prevot} for the internal reddening law. The reason 
being that in BCGs such as this, larger graphite molecules are destroyed by the
ionising radiation, producing an extinction law without any bump at 2200\AA, 
similar to that of the SMC (Mas-Hesse \& Kunth, 1999). 
By the same method as for metallicity and IMF, we investigated how the 
choice of extinction law effects our study. The effect on the pixels in
the CTN map of switching to the LMC law (Fitzpatric 1985) can be seen in
the bottom panel of figure \ref{varypara}. The Galactic law of (Cardelli 
et al, 1989) and the Calzetti law (Calzetti, 1997) 
showed a similar downward trend in CTN but to a greater extent

The plot shows that, if the LMC law were applied instead of that of the SMC,
the general trend would be to decrease the CTN factor at each pixel. 
The effect is more pronounced for larger CTN factors.
Unlike 
in the cases of metallicity and IMF, this {\em is} what we would naively 
expect. The effect seen is attributed to the fact that the LMC law provides 
less extinction than the SMC at short wavelengths. This causes the 
spectra to which the LMC law was applied to have steeper UV continua than 
the same initial spectra 
reddened with the SMC law; the result is a lower CTN factor. This 
effect is also present in F140LP-F336W colour where the points are
shifted towards the blue. The choice of extinction law has little 
impact on extinction at large wavelengths but varies by $\sim 50$\% at 
\lya. Hence, the choice of reddening law has a larger impact on the CTN than 
optical colours. In the \lya\ line-only image, the flux in the 
IUE aperture was $121\times 10^{-14}$ erg~s$^{-1}$~cm$^{-2}$ when using the
LMC law. This decrease in \lya\ flux was also present when using the
Cardelli law ($112\times 10^{-14}$ erg~s$^{-1}$~cm$^{-2}$) and the 
Calzetti law ($76\times 10^{-14}$ erg~s$^{-1}$~cm$^{-2}$).

\subsection{The stellar \lya\ feature}

There has been a long controversy over the importance of \lya\ absorption
in stellar atmospheres pertaining to observations of \lya\ from nebulae; 
eg. Valls-Gabaud (1993), Charlot \& Fall (1993). 
The synthetic 
starburst spectra comprise both stellar and nebular components with \lya\
emission from stars (as from galaxies) being subject to significant radiative
transfer effects in the atmospheres. Atmospheres themselves are based upon
empirical data that may include ISM absorption from the Milky Way in the
spectra. Moreover there is a lack of data describing atmospheres of very
low metallicity stars and there is not a single O star for which \lya\ can 
be observed due to the large ISM column. Depending on the wind properties, 
the \lya\ feature can be in absorption, a filled in P-Cygni profile or in 
emission. For the hottest stars, we expect net equivalent widths 
of only around a few \AA \ (Klein \& Castor, 1978). While the current status 
of stellar atmosphere models may be sound, atmospheres are still not 
developed enough to allow us to determine the contribution of the stellar 
\lya\ feature to overall \lya\ from galaxies.

Using the best currently available model spectra, we attempted to
investigate the effects of this feature. We adopted the latest modification
of the SED@ synthetic spectra (M. Cervi\~no, private communication) based upon
stellar evolutionary synthesis models. The stellar input spectra for these
models were built with the TLUSTY and SYNSPEC codes of Lanz \& Hubeny
(2003). These models predict Ly$\alpha$ assuming a plane-parallel geometry,
non-LTE, and full line-blanketing. The assumption of a plane-parallel
geometry is often not appropriate for Ly$\alpha$, which can form above the
photosphere in the expanding outer envelope. We expect the models to produce
too strong Ly$\alpha$ absorption, as the wind emission is neglected.
Therefore the predicted stellar Ly$\alpha$ absorption equivalent width
represents an upper limit to the observations. The spectra included stellar
and nebular continuum emission for instantaneous burst with Salpeter IMF and
solar metallicity.
The age range available was 1 to 10Myr. 
Without applying any internal reddening,
Galactic reddening or Galactic absorption to the spectra, we computed the 
CTN factor for these spectra. The stellar \lya\ feature was then removed
by interpolating linearly in between 1211 and 1221 \AA\ and the CTN factor
recomputed. We found that, at ages less than 10Myr, the removal of 
stellar \lya\ feature decreases the CTN by no more than 1.8\%.

\subsection{Stellar, nebular and emission line components}

In our analysis (section 6) of the \lya\ line-only image, diffuse emission is 
seen from regions surrounding the starbursting knots. The continuum flux in 
these regions is low and the stellar population is clearly
aged in comparison to the bright continuum sources. The H$\alpha$ 
equivalent width map (Bergvall \& \"Ostlin 2002) shows equivalent widths in
these regions of 300 to 400\AA. This stellar population is not 
capable of providing such a high $W({\rm H} \alpha)$ and the 
ionising photons most likely come from the massive young star clusters that 
inhabit the central regions.
 
The Starburst99 dataset offers a choice of stellar only, or stellar+nebular
spectra which we used to subtract out the nebular only component. We then 
used the independent components to investigate the effects of varying their
relative contributions. Such a study is of importance since the spectra
are devised for zero-dimensional objects whereas in our images, the regions
of strong nebular and ionising continua are clearly resolved. 
Boosting the nebular component has the effect of lowering the CTN 
because in the 1200-1500\AA\ region (the region of the SED from which
the CTN is  computed) the nebular continuum is always flatter 
than the stellar. 
We found that to decrease the total \lya\ flux by 10\%, we needed to boost
the nebular continuum component by a factor of 20. 
Such a boost in the 
nebular contribution would have a dramatic effect on $W({\rm H} \alpha)$ but 
the effect on the CTN remains small. 
When measuring the age in the diffuse 
\lya\ regions from our models (zero dimensional) for an instantaneous burst,
we determine ages between 15 and 40Myr. Such ages correspond to 
$W({\rm H} \alpha)$ in the range 17.5 to 0.6 \AA\ in the same model set.
Boosting the nebular 
component by a factor of 20 increases the measured $W({\rm H} \alpha)$ = 
17.5\AA\ to 350 \AA\ -- consistent with the $W({\rm H} \alpha)$ map. 
For ages found in these regions, the 
nebular continuum in the region where the CTN is computed is around 
three orders of magnitude smaller than the stellar component; hence boosting 
the nebular component by reasonably large factors makes little difference to 
the CTN. Varying
the relative contribution of the components has no effect on the \lya\ map 
that is distinguishable by eye and has a negligible impact
on the overall \lya\ flux. 

The Starburst99 dataset also offer the option of continuous star-formation as
opposed to the instantaneous burst that we had assumed initially. While we 
believe the instantaneous burst to be a good approximation in the region of 
the current
starburst, this may not necessarily be so for the underlying population.
The underlying population may contribute more strongly to the continuum in the 
regions from which we see diffuse \lya\ emission. 
The continuous star-formation model also maintains a high $W({\rm H} \alpha)$
for a much longer period, ($W({\rm H} \alpha) > 300$\AA\ at ages $> 400$Myr). 
Using the continuous star-formation model, we 
measured a \lya\ flux decrease in the aperture corresponding to that
of the IUE from
134 to 120 $\times 10^{-14}$ erg~s$^{-1}$~cm$^{-2}$. While it may fit the 
data rather well, we do not believe the continuous star-formation model
on the grounds that clusters in these regions are no older than $\sim40$Myr 
(\"Ostlin et al. 2002).


Where the starburst is young, nebular emission lines may strongly affect
the luminosity of a galaxy in certain wavebands. Hence the optical colours
may be determined by the presence of nebular emission lines, not simply 
the continuum as previously assumed.
The WFPC2 images cover a large fraction of the optical domain. 
From a galaxy such as ESO\,338-IG04 we can expect strong emission lines which 
may fall within our filters (eg. the F555W filter transmits the strong
O[{\sc iii}] line). Since the model spectra we used do not include nebular
line emission, we investigated how our results
may, or may not be influenced by the addition of lines to the spectra. 

The Mappings III photoionisation code of Kewley et al (2004, in prep) was 
used to model nebular emission lines for a number of different burst ages
via the Starburst99 interface. CTN factors and optical colours for all our 
filter possibilities were computed with and without the inclusion of 
nebular emission lines. The \lya\ line was not included since the definition
of CTN requires its absence.

For a burst of 5Myr, the nebular emission lines did not affect the 
CTN by more than 1\permil\ or either of our selected colours by more than 0.05 mag. 
Nebular line emission was therefore not considered important for the study
of this galaxy with the data that we have available.

\subsection{Possible departure from a single stellar population}

We also examined how sensitive the results are to the assumption of a 
single stellar population. Our concern originated from the possibility 
that the underlying stellar population may contaminate the optical colours
of the pure starburst. We assumed that the old population may contribute 
10\% of the flux in the B-band and used the 900Myr Starburst99 spectrum 
as a template for this old population.
For each age, we computed a new spectrum comprising the standard spectrum
for that age, plus that of the old population normalised to 10\% of the flux of 
the standard spectrum at 4000\AA. Again this modification produced a \lya\
line-only image that was visually indistinguishable from that produced with
the standard model parameters. In the IUE aperture, this modification
produced a \lya\ flux of 122 $\times 10^{-14}$ erg~s$^{-1}$~cm$^{-2}$; a 
decrease of 9\%. 

Due to the very low UV flux of the old population, it has no discernible
effect on the CTN and acts only to redden the computed optical colours.
The effect was most noticeable for young ages in the F336W-F439W 
colour where it amounted to an increase of $\sim 0.1$ mag for a 1Myr 
unreddened burst. The effect of the underlying population was considered not
to be an issue in the analysis of ESO\,338-IG04.

\section{Analysis and Discussion}

The two-colour technique described in section 3.1 proves an excellent
method of subtracting the continuum from the line in our ACS imaging 
survey. 
Disentangling the degenerate effects of age and reddening relies upon
the sampling of the continuum slope in critical regions. In our technique we
use one colour that is very sensitive to the reddening (F140LP-F336W, 
sampling the UV continuum slope \bet) and one colour that is sensitive to
age (F336W-F439W, sampling the 4000\AA\ discontinuity).
Not only does the qualitative comparison of the continuum subtracted
image compare very well with the STIS spectra along the slit, but 
quantitative photometric results compare well with previous values of the 
\lya \ flux and equivalent width. 

In addition, an important point is that our results are not 
highly model dependent. 
Qualitatively, the line-only \lya\ images were indistinguishable from
one another; each showing the same features of absorption and emission 
and each showing the diffuse emission regions. Moreover, when comparing 
fluxes, all were in agreement to better than 50\%, even when parameter space
well outside of the known constraints was investigated (eg, 
IMF $\alpha$ = -3.3 or twice solar metallicity).

ESO\,338-IG04 is a well studied galaxy and this data can be compared with
a wide array of other studies, against which our results can be tested.

\subsection{The \lya\ emission}

The false-colour map, (figure \ref{colew}) 
shows how \lya, UV continuum and [O{\sc iii}] compare spatially. 
[O{\sc iii}] emission can be used to gain insight into hardness of the
ionising flux since it maps twice ionised oxygen. 

There are several noteworthy features in this image.
The hot young star clusters, very bright in the UV continuum appear as 
the bright green blobs. These are the regions where the continuum
flux is high and the slope is steep due to the young stellar population. 
This is the reason why
so many of the star clusters appear in green. All of the regions 
surrounding the green and yellow star clusters show [O{\sc iii}] emission 
to some degree. Here the UV flux from the cluster is still strong enough 
to twice ionise oxygen at large distances from the UV signature of the 
cluster. This effect can be seen very strongly around knots B and F.

Knot A is another interesting case. Nearly the whole region around knot A 
shows up in either cyan or green, indicating, not only a strong UV
continuum but also a positive flux of \lya. This is consistent with the
STIS spectrum of this knot. \lya \ escape from this region due to the 
cluster driving an outflow of the local neutral ISM,  
giving rise to the P~Cygni
profile that can be seen around 1228\AA\ in figure \ref{voigt}. In the 
spectrum, \lya\ can be seen in absorption and emission, bluewards and 
redwards of 1228\AA\ respectively. Just to the
west of knot A, the cyan (\lya\ and UV) becomes yellow (UV and O[{\sc iii}])
as the \lya\ flux falls off but oxygen is still twice ionised. 

Diffuse emission regions are very evident surrounding the starburst regions.
In these regions the continuum flux is very low and \lya\ equivalent
widths exceed 200\AA. We explain this find as \lya\ photons that escape the 
H{\sc i} region either by diffusing to a region where the optical depth to
infinity is less than unity or by scattering into the wing of the 
emissivity profile. These diffuse emission regions are of great importance
for the overall line luminosity. As can be seen from the surface brightness 
profile, these regions dominate the luminosity output and the scattering
mechanism is responsible for the majority of the \lya\ escape.

Analysis of the strong absorbing regions shows the \lya\ equivalent width to
be around -65\AA, indicating damped absorption due to a static covering of 
neutral hydrogen. This value for the equivalent width in absorption compares 
well with estimates of equivalent widths of \lya\ absorption in other BCGs. 
Cursory inspection the GHRS spectra of SBS\,0335-052 presented in 
Thuan \& Izotov (1997) and of I\,Zw 18 presented in Kunth et al (1994) show
\lya\ equivalent widths of $\lesssim-50$\AA\ and $\lesssim-45$\AA\ 
respectively. 

\subsection{\lya\ correlations with other properties}

\"Ostlin et al. \cite{ostlin03} mapped the age of the star clusters in this 
galaxy, the youngest having ages $\lesssim2$Myr. 
These are the regions from which
we expect \lya \ to be trapped if the local ISM is not configured so as to 
allow its escape. Towards the west of the galaxy, this age map shows many 
massive, very young clusters which coincide exactly with the absorption from knots
E, B and the surrounding regions. Other young clusters are marked in \"Ostlin et 
al. to the south and south-east of the central cluster. These also appear as 
absorption features in our continuum subtraction. 

We compared the current results with the results from \"Ostlin et al. 
on the age distribution of compact star clusters. We selected the young
($<30$ Myr) sub-population and looked for systematic relationships 
between their properties
(age, photometric mass, luminosity, colour, radial distance from the
centre) and the local \Lya\ emission or absorption fluxes and equivalent 
widths. We note that we are comparing regions which are coincident 
on the 2D projected image, and that are not necessarily spatially 
coincident in 3 dimensions.

Tenorio-Tagle et al (1999) and Mas-Hesse et al (2003) suggested that
\lya\ emission from  starburst is expected to be a function of evolutionary
and geometrical effects. In this model, pure or P-Cygni \lya\ emission
can be expected only if we are looking into the ionised cone. 
No \lya\ should be 
expected along lateral sightlines that view the surrounding H{\sc i}.
We find no correlation between the \lya\ escape/absorption or equivalent width
and age of the comparison cluster. 
Hence we are not currently finding any support for the evolutionary model.
We stress, however, that \lya\ escape is convolution of many
parameters, not simply cluster age, and geometry may provide
the scatter that masks the trend. 

\begin{figure}
\resizebox{0.93\hsize}{!}{\rotatebox{-90}{\includegraphics{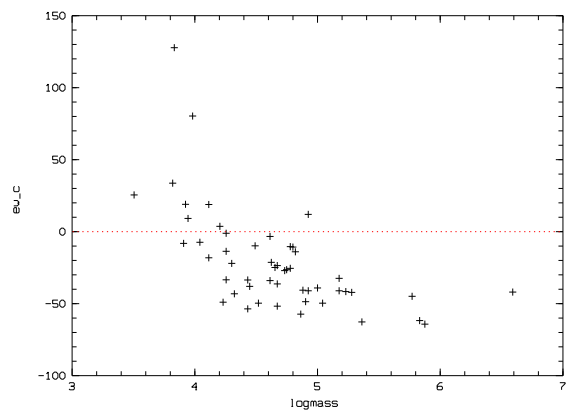}}}
\caption{Equivalent width of \lya\ in \AA\ for a $3 \times 3$ pixel 
box around young clusters ($<30$Myr) vs. log cluster mass in $M_{\odot}$.}
\label{ewvslogmass} 
\end{figure}

The only clear trend is seen when comparing equivalent width and luminosity
or mass: the more luminous (or massive) the cluster, 
the smaller the EW is on average (and in fact it is predominantly negative),
see figure \ref{ewvslogmass}.
This serves to reinforce the find that \Lya\ escape is not 
regulated by the output
UV luminosity from the cluster. This trend is expected if one assumes that
the more massive clusters are associated with more gas, hence a larger
neutral hydrogen column density.

\subsection{\lya\ and other kinematical tracers}

The ionised gas kinematics of this galaxy have been explored through 
Fabry-Perot interferometry, described in \"Ostlin et al. (1999, 2001). 
This study revealed a complex and perturbed, almost chaotic velocity field.
Photometric studies have shown the tail to the eastern
side to contain a distinctly different stellar population from the rest.
The authors explained this as the result of a merger which was the 
triggering mechanism for the current starburst.
The merger may be responsible for triggering the starburst, producing the
\lya\ photons and is also responsible for the irregular ISM kinematics 
that allow these locally produced photons to leave the galaxy. 
One striking result of this study was the discovery that in the
region of diffuse \lya\ emission north of knot A, the ionised gas is 
redshifted with respect to the stellar population. 
Regions south of knot A are similarly blueshifted. This could either
point to a a bipolar outflow of the ISM or the presence of a second
kinematical component with a centre very close to knot A and kinematic 
axis in the north-south direction. This axis has the same orientation
as the one we find for diffuse \lya\ emission.


A UVES spectrum of knot A discussed
in \"Ostlin et al. (2005 in prep) shows that the ionised gas around this
knot has at least two components: one which is redshifted by $\sim$25 
km~s$^{-1}$
and one blueshifted by $\sim$40 km~s$^{-1}$. In addition, Na{\sc i} D
absorption lines which probably probe the neutral ISM, are blueshifted 
by $\sim$20 km~s$^{-1}$ compared with the stellar component. This suggests
that gas is flowing towards us along the line of sight and is
consistent with the P-Cygni interpretation of the \lya \ 
emission around knot A. South of knot A the UVES spectrum reveals 
an ionised component blueshifted relative to knot A but also components 
redshifted by 100 to 200km~s$^{-1}$, showing the complex nature of the medium.

The VLA H{\sc i} mapping of ESO\,338-IG04 and its companion also suggests a 
recent strong interaction between the two galaxies (Cannon et al 2004).
Although
the H{\sc i} map has a much lower resolution than the ACS images, it is
suggestive that the regions with strong diffuse \lya \ emission  north and
south of knot A coincide with a region where the  H{\sc i} column density
is smaller (N(H{\sc i}) $< 5\times 10^{21}$ cm$^{-2}$) than elsewhere in
the galaxy. The other strong diffuse
\lya \ region to the west of knot A has 
N(H{\sc i}) $ > 7\times 10^{21}$ cm$^{-2}$. In the region near knot A, 
although we stress again that the resolution makes a comparison with a
point source even more uncertain, the column density is
N(H{\sc i}) $\approx 6\times 10^{21}$ cm$^{-2}$.

All these column densities would be high enough to produce damped 
absorption of \lya; something we do see in parts of the central starburst, 
despite its extremely low dust content. That any emission at all 
comes from the central UV-bright knots must be attributed to a 
clumpy ISM and to velocity shifts of neutral gas relative to the photon
producing regions. Note however that velocity shifts of H{\sc i} (as well as 
H$\alpha$ discussed above) over the galaxy are relatively small.

Synthesis of these kinematical tracers leaves an incomplete picture.
The Fabry-Perot and UVES spectra hint that in the regions
of diffuse \lya\ emission, the ISM may be subject to a rather slow 
large-scale outflow
of gas (and hence allowing the escape of \lya\ photons) but also reveal
a highly perturbed medium. 
Conversely, the H{\sc i} data shows the neutral ISM to be subject only to
very small velocity shifts over the whole galaxy (although the resolution
is much poorer than the scales explored here), indicating the diffuse 
\lya\ emission may be the result of a clumpy medium, perturbed on a much 
smaller scale.

\subsection{The diffuse emission}

It has been shown in section 5 that our results are not highly dependent upon the 
model parameters chosen. One striking result of the photometric study was that 
over 70\% of the total \lya\ originates from the diffuse emission regions. 

Our unreddened  \lya \ flux in the IUE aperture amounts to 
$134 \times 10^{-14}$~erg~s$^{-1}$~cm$^{-2}$. However,
this comprises a combination of positive and negative contributions and
the net contribution from the central burst region is negative.
We can divide the total \lya \ flux into 3 spatial components:
central regions with positive \lya \ flux, central regions with negative
\lya \ flux, and the region outside, i.e. where we see diffuse emission:
  
\begin{table}[h]
\caption[]{\lya\ Fluxes and equivalent widths from the various components} 
\begin{flushleft}
\begin{tabular}{lll} 
\hline
\noalign{\smallskip}
Region             &     Sum of flux                &     EW      \cr
                   &     ($10^{-14}$~erg~s$^{-1}$~cm$^{-2}$)   &     (\AA )   \cr
\hline
\hline
Centre positive    &    93                          &    47   \cr
Centre negative    &     -163                       &     -78   \cr
\hline
Centre total       &     -70                        &     -17   \cr
Diffuse            &     204                        &     113   \cr
\hline
Total              &     134                        &     22.6       \cr
  
\hline
\end{tabular}\\
Notes:\\
Values in this table refer only to fluxes in the IUE aperture. 
\end{flushleft}
\label{tablyacompontents}
\end{table}

Had it not been for the positive diffuse emission, this galaxy
would not have any \lya \ detected by the IUE satellite.

We compare these numbers with the  H$\alpha$ flux, obtained
from ground based narrowband imaging (Bergvall \& \"Ostlin). Of the total
H$\alpha$ flux ($f_{\rm H\alpha}=290 \times 10^{-14}$~erg~s$^{-1}$~cm$^{-2}$,
 \"Ostlin et al. 
2001, Gil de Paz et al. 2003), 87\% comes from the region covered by our
\lya \ observations. Of the H$\alpha$ flux in this region approximately 35\%  comes
from the diffuse region and the rest from the central regions.
The uncertainty
on these quantities is about $\pm$5\% and arises due to the very different
spatial
resolution of the HST/\lya \ and the ground based H$\alpha$ images. Of
course we do not have the resolution to differentiate the H$\alpha$ flux
of the regions showing positive and negative \lya \ flux.

Correcting for Galactic extinction, \lya /H$\alpha = 3.3$ for the diffuse
region. The  \lya \ photons in this region could either be locally
produced
by ionising flux from the central burst (as there are no UV-bright sources
in the diffuse region, see \"Ostlin et al. 2003), or be due to \lya \
photons
diffusing out from the centre after multiple scatterings. The former
alternative
appears more realistic since we see H$\alpha$, which is not subject to
multiple scattering, in the diffuse region and moreover, the H{\sc i} column
density in the diffuse region is high and with little velocity difference
with respect to the central burst (Cannon et al. 2004). If we assume
that the internal reddening in the diffuse region is similar to that in
the
centre ($E(B-V)=0.05$, see also \"Ostlin et al. 2003) and assume the SMC
reddening law, this would lead to a corrected \lya /H$\alpha$ ratio of
6.6, or about half the value predicted from case A recombination. To
reconcile these numbers, either we have to assume that the internal
reddening in the diffuse region is a factor of $>2$ higher than in the
centre, or that each \lya \ photon has suffered, on average, $\gtrsim2$
additional
scatterings whereby the total \lya \ intensity has been reduced by the 2-photon
process. The latter appears very reasonable given the large neutral
hydrogen column density. Hence, the diffuse emission is best explained
by ultraviolet photons escaping the central burst through a porous medium
and ionising hydrogen in the diffuse regions, after which \lya \
photons are scattered a few times before escaping.

In the central region, the absorption dominates over emission. The
predicted
equivalent width of \lya \, disregarding any absorption processes, is on
the order of 200 \AA, whereas damped absorption may lead to negative
values
$< -50$\AA. Hence, to create a net absorption more than 80\% of the
photons
with wavelength near $\lambda_{\rm em}=1216$ \AA\ have to be absorbed by
static (or nearly static) neutral hydrogen. That we see any pixels with
positive \lya \ must be attributed to the porosity of the ISM and local
velocity differences between the ionised gas and neutral medium in front
of it along the line of sight. Looking at the central H$\alpha$ flux and
attenuating it by Galactic reddening and an additional internal reddening
of $E(B-V)=0.05$ using the SMC law, we would expect to see a flux of
approximately  $1000 \times 10^{-14}$erg~s$^{-1}$~cm$^{-2}$ if H{\sc i} absorption was
neglected. The central \lya \ photons that do reach us amount to
$93 \times 10^{-14}$erg~s$^{-1}$~cm$^{-2}$, or about 9\%. Hence, this is consistent
with more than $80$ \% of the photons being destroyed. The pixel scale
of ACS/SBC corresponds to a linear projected
scale of 5pc at the distance of 37 Mpc.
Possibly the number of positive central \Lya\ photons would be higher with
even better spatial resolution.

Taking the H$\alpha$ flux from the area corresponding to the ACS images,
we corrected this flux for Milky Way reddening and for internal reddening
using the SMC law with $E(B-V)=0.05$. Assuming case A recombination, we 
computed the total \lya\ flux produced to be 
$3860 \times 10^{-14}$erg~s$^{-1}$~cm$^{-2}$. 
By comparing this value to the  detected \lya\ flux (corrected for 
Galactic extinction) of 
$194 \times 10^{-14}$erg~s$^{-1}$~cm$^{-2}$ 
we obtain an escape fraction of 5\%. This escape fraction is an 
order of magnitude lower than what one would derive  (44\%)
by assuming internal reddening only.

\subsection{Implications for high-{\em z} surveys}

Our technique of performing a continuum subtraction
is dependent upon careful modeling of the SED. We notice that the region
of the SED imaged by our offline filter (central wavelength: 1400\AA;
rectangular width: 250\AA) corresponds almost exactly to the region of the
SED of a galaxy a redshift of around 4, imaged by a broadband Johnson R
filter. Malhotra \& Rhoads (2002) used just this broadband filter to subtract
continuum from their sample of galaxies in the redshift range 4.37 to 4.57 using
a variety of online narrowband filters. Hence, their continuum filter maps
a very similar part of the restframe SED to ours. 
The \lya\ galaxies in their sample
were found to have higher than expected equivalent widths with
a median value of 400 \AA. 
It is therefore interesting to point out that,
when we assumed a flat continuum we derived a \lya\ flux for
ESO\,338-IG04 of 430 erg~s$^{-1}$~cm$^{-2}$ and equivalent width of
72\AA -- over 3 times higher than the values obtained through SED
modeling. 
Making naive assumptions about the continuum slope caused us to vastly
overestimate the \lya\ equivalent width. To perform accurate high-$z$
surveys, we suggest it may be necessary to at least employ more offline
filters so as to accurately model the continuum.

\section{Conclusions}

We have presented high resolution \lya \ imaging of the local starburst galaxy
ESO\,338-IG04 based on observations with the HST and the solar blind channel of
ACS, complemented by available WFPC2 data. We show that a careful continuum 
subtraction is crucial and have demonstrated that we have a technique in place to 
subtract the continuum provided the necessary supporting data is available.
In our analysis of Starburst99 synthetic spectra, we were able to 
disentangle the effects of age and internal reddening by carefully modeling
the evolution of the UV continuum slope and 4000\AA\ discontinuity.
We present a total \lya\ line flux of 
$f_{\rm Ly\alpha,TOT} = 194 \times 10^{-14}$erg~s$^{-1}$~cm$^{-2}$
and demonstrate the photometric validity of our technique by comparison with
STIS spectroscopy and  the 
IUE data presented in Giavalisco et al.(1996). Our  \lya\ flux of 
$134\times 10^{-14}$ erg~s$^{-1}$~cm$^{-2}$ matches that obtained with the IUE
satellite of $123\times 10^{-14}$ erg~s$^{-1}$~cm$^{-2}$. Within the errors 
associated with the IUE, these values are in perfect agreement.  
We also find a total \lya\ equivalent width of 22.6\AA, again in agreement with the value 
obtained with the IUE. 
We find extensive evidence for resonant photon scattering of \lya\ in diffuse 
\lya \ emission regions that surround the central starburst. 
These regions dominate the total output
luminosity of the line and equivalent widths in here exceed 200\AA. 
If it would not have been for the diffuse emission, this galaxy would 
not had been seen as a net emitter of \lya \ e.g. in IUE spectra. 
By comparison of the \lya\ and H$\alpha$ fluxes in these regions, we 
calculate that each detected \lya\ photon has suffered
2 additional scatterings, 
reducing the net \lya\ emission via two-photon emission. 
We estimate that of all the \lya\ photons produced, only 5\%  leave the
galaxy. We find
that the \lya\ line traces the kinematics and perturbed morphology of 
ESO\,338-IG04, correlating closely with other tracers of kinematics.
In all, this galaxy provides an excellent laboratory for studying the
physics of \lya\ emission, displaying a variety of mechanisms:
emission through P-Cygni features from the central knot (A), damped
absorption, leakage of central  \lya\ photons through paths with
small optical depth, and diffuse emission.
While our technique is entirely reliant upon the use
of synthetic spectra, it is not highly dependent upon the choice 
of parameters (metallicity, IMF, etc.) nor on the choice of internal reddening law. 
We discuss the implications for the understanding of \lya\ escape from galaxies
and the nature of \lya\ emitters at high redshifts, and find that if careful 
attention is not paid to the slope of the continuum, then fluxes and equivalent 
widths may be overestimated by a factors of $\sim3$.

\begin{acknowledgements}
MH and G\"O acknowledge the support of the Swedish National Space Board
(SNSB) and the Swedish Research Council (VR).
JMMH was supported by Spanish MEC under grants AYA2001-3939-C03-02 and         
AYA2004-08260-C03-03.
We thank M. Cervi\~no for his help with the high resolution continuum models 
and J.~M. Cannon \& E.~D. Skillman for their contribution to this study.
This work was supported by HST grant GO-9470 from the Space Telescope
Science Institute, which is operated by the Association of Universities for
Research in Astronomy, Inc., under NASA contract NAS5-26555.
\end{acknowledgements}

{}

\end{document}